\documentclass[12pt]{article}

\usepackage{mathrsfs}
\usepackage{epic, eepic}
\usepackage{subfigure}
\usepackage{amsfonts, amssymb, amsmath, graphicx, epsfig, lscape, capt-of}
\usepackage{url}
\usepackage{ifpdf}
\usepackage[boxed]{algorithm2e}
\usepackage{array}
\usepackage{booktabs}
\usepackage{multirow}
\usepackage{xcolor}
\usepackage{adjustbox}

\def\inst#1{$^{#1}$}

\makeatletter

\usepackage[normalem]{ulem}

\begin{document}%


\title{
Economic Freedom :\\
the top, the bottom, and the reality. \\
I. 1997-2007  \\
}
 
\author{%
Marcel Ausloos\inst{1,2,3,*}, \\ Ph. Bronlet.\inst{3,4}
  }
 
\maketitle

\begin{center}
{\footnotesize
\inst{1} 
 School of Business, University of
Leicester,  Brookfield,  Leicester, LE2 1RQ, UK\\
\texttt{ma683@le.ac.uk}\\
\vspace{0.3cm} \inst{2} Department of Statistics and Econometrics, 
Bucharest University of Economic Studies\\
6 Piata Romana, 1st district, Bucharest, 010374 Romania\\
\texttt{marcel.ausloos@ase.ro}\\
\vspace{0.3cm} \inst{3} GRAPES,
Sart Tilman, B-4031 Liege, Belgium\\
\texttt{marcel.ausloos@ulg.ac.be}\\

\vspace{0.3cm} \inst{4}
AMOS, 
Rue des Chasseurs Ardennais 2, B-4031 Liege, Belgium\\
\texttt{p.bronlet@gmail.com}\\

\vspace{0.3cm} \inst{*} corresponding author: marcel.ausloos@uliege.be
}
 \vspace{0.35cm}
 \vspace{0.35cm}
\end{center}
\newpage

\begin{abstract}
 
 We recall the historically admitted prerequisites of Economic Freedom (EF). We have examined 908 data points for the Economic Freedom of the World (EFW)  index and 1884 points  for the Index of Economic Freedom (IEF); the studied periods are 2000-2006 and 1997-2007, respectively, thereby following the Berlin wall collapse, and including  Sept. 11, 2001.  After discussing EFW index and IEF, in order to compare the indices, one needs to study their overlap in time and space.  That leaves 138 countries to be examined  over a period extending from 2000 to 2006, thus 2 sets of 862 data points. The data analysis pertains to the rank-size law technique. It is examined whether the distributions obey an exponential or a power law. A correlation with the country Gross Domestic Product (GDP), an admittedly major determinant of EF,  follows,  distinguishing regional aspects, i.e. defining 6 continents. Semi-log plots show that the EFW-rank relationship is exponential for countries of high rank ($\ge  20$);  overall the log-log plots point to a behaviour close to a power law. In contrast, for the IEF, the  overall ranking has   an exponential behaviour;  but the log-log plots point to  the existence of a transitional point between two different power laws, i.e., near rank 10. Moreover,   log-log plots of the EFW index relationship to  country GDP   is characterised by a power law, with a rather stable exponent ($\gamma \simeq 0.674$) as a function of time. In contrast,   log-log plots of the IEF relationship with  the country's gross domestic product  point to a downward evolutive  power law as a function of time. Markedly the two studied indices provide different aspects of EF.

\end{abstract}
 \vskip5cm
 
\clearpage

\section{Introduction}\label{Introduction}

Numerous empirical studies  \cite{[1]} pretend to show that Economic Freedom (EF) favours economic growth, prosperity, poverty reduction and has many other beneficial effects, beside being also a condition necessary for the development of democracy.   However, before proposing   modern theories of Economic Freedom, it seems that one should   first wonder about  EF definition, and have proofs that Economic Freedom exists. The goal of this paper is to study the world EF situation before the recent  (XXI-th century) economic crisis. A second paper is intended for later years as explained below. In brief, this is due to  different definitions and changes in geo-political economic conditions. It is expected that the paper can be useful for econo-physicists and other researchers, due to the somewhat original approach, more numerical, i.e. along the lines of econophysics thought. 

The oldest of these publications, {\it The Wealth of Nations}  by Adam Smith in 1776, shows that the preservation of  individual freedom  to pursue their own interests is due to the necessity of creating   a social and  more prosperous civilisation  \cite{Smith}.
On the other hand, protectionism and trade performed under a monopoly (like that of the British empire at the time of Adam Smith) have the necessity to preserve the status quo and to privilege a few elites. Frederic Bastiat shows, in {\it  Economics Harmonies} \cite{Bastiat64}, that all human actions lead to care and harmony if these actions are motivated by private considerations. Thus, Bastiat recommends, or even advocates, "liberty"\cite{Bastiat44},; in our own words,  EF  contains so much creativity that it leads to many opportunities for bettering human life.

But what is "economic freedom" (EF) ? A simple definition   among many similarly proposed by others  may be as follows:
The freedom of the economy is the freedom to produce, exchange and consume any goods and services acquired without use force, fraud or theft.

In order to have a more complete appraisal of EF, one might consider James Gwartney and Robert Lawson article  \cite{[2]}. Gwartney and Lawson do not give a proper term  for   economic freedom, but claim to provide all the conditions to be met in order to obtain "economic freedom":  in brief, the foundations of any "economic freedom"  is the respect of the "rule of law", of property and privacy,  i.e., "right to own", and demands freedom for agents wishing to enter into contracts, i.e. "freedom to contract". Thus, before, measuring EF and discussing such  measures, let us briefly examine the framework in the following three subsections.

\subsection{Rule of Law}\label{ruleoflaw}
Many theoreticians of the economic liberalism maintain that  the aim of the prerequisites for EF is the establishment of a rule of law; e.g. \cite{GollwitzerIMF}.
A "Rule of Law" ({\it "Etat de droit"}) is an institutional system in which the government and the individuals are subject to the law. This right shall apply in an identical way to each individual and to all economic agents.

This principle of equality of individuals before the law is the guarantee that the  fundamental rights of   citizens will not be raped by those in power.  It also excludes any form of privilege, i.e. the application of the law  with  the purpose of favouring one group of people over another. It restricts also any arbitrary application of the law.  Otherwise, one of these "misactions" would lead to a restriction of economic freedom.

\subsection{The right to own}  \label{righttoown}
The second prerequisite for EF is the respect of the individual rights to own property. To achieve this, a system must be established which ensures the right to use ($usus$) and to profit ($fructus$)  from this property. The system shall also ensure the right to transfer this property  to another person  as long as they are both consenting.

These fundamental rights are the guarantees that individuals will be able to be autonomous and   will have the opportunity to seek to achieve their own goals.
Many economists, such as Milton Friedman  \cite{[3],[3b]} or Murray Rothbard  \cite{[4a],[4b]}, consider the right of ownership  as the most fundamental of the rights, of all other rights. It guarantees individuals to have  individual freedom  and allows for better personal development than otherwise, under a regime of coercion. It also reduces uncertainty and encourages investment by creating favourable conditions for an economic development.

Empirical studies  \cite{[5]} show that countries with a right to own   have an economic growth rate almost twice larger than countries where this right is not respected. 
According to (the Peruvian economist) Hernando de Soto  \cite{[6]}, a large part of the poverty in third world countries is caused by the system  lack of favouring some equality and by the absence of a right of ownership.  

\subsection{Freedom to contract} \label{righttocontract}

A contract is an agreement between two or more parties, having  the purpose of establishing obligations  at the expense of each of those parts. The freedom to  contract  contains therefore the right to choose the  parties with which the contract is formed and to agree on the content of this contract (what to give, to do or not to do). The parties have the right to choose the subject of the contract, but once the contract has been made, they are obliged to fulfil  the terms of the contract.

The main economic function of contracts is to transfer rights of one individual's property to another person.

\subsection{Other  definitions of economic freedom}

The  Gwartney and Lawson definition  \cite{[2]}  is an ideal one, but accepted by the classic liberal economists. It is intimately linked to a respect of the law which in so doing   protects individuals against external aggression that would   aim to take ownership of their  property. This  definition is valid only in a "non-negative legal context".

There are many other  definitions   of EF but none is unanimously accepted.

Examples of "economic freedom" in a "positive law context" are given  by Amartya Sen  \cite{[7]};
Amartya Sen argues for an understanding of freedom in terms of capacity of an individual to achieve his/her own goals. Notice that in a similar line of thought, Goodin, Rice, Parpo, and Eriksson    \cite{[8]} propose to measure  "freedom", even outside financial or economy considerations,  from the  available time that   people have in participating in  an  activity  so chosen by them.

\subsection{Paper content }
However, before a theory of Economic Freedom is proposed, should one not first have proofs that Economic Freedom exists, - where ?, when ?
In fact, the questions demand a study of other highly fundamental research questions, in particular about the measurement(s) of Economic Freedom(s?) themselves, and on the meaning of the measures (so called "indices"). Immediately tied to the former and the latter,  the correlations with other socio-economic measures should be considered in order to provide stylised data for some preparation of modelling, later on with determinants or/and components. These are huge challenges having led to a  vast literature.

Thus, even though the literature is enormous, on many aspects, we have only considered some, in our opinion, very elementary but fundamental,  ground level basis, accepting two types of measures, explicitly defined  in Section \ref{EconomicFreedomIndices}: the Economic Freedom of the World (EFW)  index and   the Index of Economic Freedom (IEF). We have examined 908 data points for the  EFW   index and 1884 points  for the  IEF; the studied periods cover  2000-2006 and 1997-2007, respectively, thereby following the Nov. 09, 1989 Berlin wall collapse and including  Sept. 11, 2001. Notice that we presently exclude the 2008 financial crisis, and the following years, due to recent economic, geopolitical, changes, and because a new definition of the IEF was recently implemented. Some further work is intended over the more recent period  (to be paper II.) in order to provide a complementary analysis later; paper II. will also contrast the findings, whence prompting any dynamical aspect.

 In order to compare the indices, one needs to study their overlap in time and space.  That leaves 138 countries to be examined  over a period extending from 2000 to 2006, thus 2 sets of 862 data points.
Since each country presents a combination of freedoms, and restrictions to freedoms, it is of interest to observe whether the country ranking contains or hides such a variety of dimensions. Due to the aimed scope of this paper, we will only care about the most often admitted primary determinant of a country economic growth (EG), i.e. the  country Gross Domestic Product (GDP).
 
 Thus, our data analysis pertains to the rank-size law technique. It is going to be examined whether the
 measures of EF have a statistical distribution which   
 follows either an exponential or a power law.    This is a sort of research question not considered  in economics classical realms, but should be of interest in econophysics.   A correlation with the country's gross domestic product (GDP)   follows,  distinguishing regional aspects, i.e. defining 6 continents.

The table of content of this paper may be as follows:
 
 In Section \ref{EconomicFreedomIndices}, we recall the  definition and content of the Economic Freedom of the World (EFW) Index and the  Index of Economic Freedom (IEF), respectively.
 
 In Section \ref{Data},  we present the extracted data, i.e. 908 data points for the EFW index and 1883 for the IEF  on the studied periods, 2000-2006 and 1997-2007, respectively.
 
 In Section \ref{relationships},  we provide the empirical laws, on one hand, the rank-size laws for both indices, plus, on the other hand, the (regression) relationship between such indices and the gross domestic product of the countries of interest. We also provide a study of regional aspects through a grouping of countries according to their geographic positions.
 
 In Section \ref{Conclusions}, we provide conclusions pointing to the weak evolution of indices over the considered time interval. We suggest lines for further research.

\section{
Economic Freedom Indices}\label{EconomicFreedomIndices}

 We position our paper within the  scholarly contributions  having  investigated, on one hand "measures of Economic Freedom", in  modern times, and 
the link between EF and  EG. Our article explores this possibility by means of a regional analysis, which we conduct on two indicators. Let us summarise the literature from such points of view.
 
 \subsection{ Economic measures}

\subsubsection{
Economic Freedom of the World (EFW) Index}

The Economic Freedom of the World (EFW) Index, published by the Fraser Institute  \cite{[18]}, is the result of a project  spanning
  20 years. It was developed after a set of conferences given by Milton Friedman and Michael Walker between 1986 and 1994, in a project gathering more than
60 of the greatest economists of the time \cite{[14]}. The aim was to create a ("strong")  base with quantifiable and objective data following a transparent procedure. Thus, anyone  could use the index  whatever  its goals and political ideals.

The EFW index measures the degree of economic freedom in 5 major "areas":
\begin{itemize}
\item 1. The size of the government, that means, public expenditure, taxes, influence on the economy
\item 2. The legal structure  
which guarantees the right to own  
\item 3. The access to a healthy currency
\item 4.   Freedom in international trade
\item 5. Regulation of costs, work and economy
\end{itemize}

For each of these 5 domains, several variables are measured, resulting in  a set of 21 components included in the index. Each component is placed on a scale going from 0 to 10.   The value 0  refers to zero freedom while the value 10 represents total freedom. Once these  components are quantified, they are averaged in order to obtain the index value. 

Several methods have  been studied for doing  such an average: without being exhaustive, one considers the weight equivalent to each component;  another  
  gives  an inversely proportional weight to the standard error of the distribution of the component, - in the various studied countries.    A third  method calls upon a panel of economists who estimate the weight that each component must  have;  the final weight being the average weight obtained from the panel members appraisals. 
A fourth method  uses the primary component analysis  
 technique to determine each weight. This latter method has the advantage   of reducing the importance of anomalies (outliers) in  estimating correlations between the components.

 Since none of these methods is really satisfactory (from our investigations, the index does not seem to be very sensitive to changes in weight), the weight  choice is not further discussed, and taken as the most simple one. 
Thus, an equal weight  for each component  is chosen in the forthcoming analysis here below. The index, so constructed, provides a value between 0 and 10 for each country.  A country with an index value close to 10 is a country where "economic freedom"  is "very large". A country with a value close to 0 is a country where EF is "non-existent".

Of course, it is expected that  each country presents a ''combination of freedoms''.  
Recently,  Lawson et al. \cite{lawson2020determinants}  
have reviewed  the determinants of  EF, with a time dependent point of view.  
  Some of the most consistent findings  
  are that current levels of EF are strongly correlated with past levels. Lawson et al. deduce that freer countries have more difficulty continuing to improve their economic freedom.

\subsubsection{
  Index of Economic Freedom (IEF)}
  
Another measure of economic freedom, published by the  Heritage Foundation \cite{[16]} and the Wall Street Journal \cite{[17]}, is the Index of Economic Freedom (IEF), - which was initiated in 1995 \cite{[21]}.  

The index was built on a set of 10 specific components    \cite{[22]}:

\begin{itemize}
\item 1. Tax freedom: it measures the importance of fiscal fees by the government on the income of individuals but also that of businesses. 
\item 2. Government  spending: it measures the total government spending. 
\item 3. Free trade: it measures the absence of commercial   barrier, affects the import and export of goods or services. 
\item 4. Investment  freedom: it measures the freedom of capital flows. 
\item 5. Financial freedom: measurement of the credit system and also of the banking system of its independence from the government. 
\item 6. Property rights: they are measures of the ease with which individuals acquire a property of their own. 
\item 7. Corruption: it measures the importance of corruption in the economic world. 
\item 8. Business undertaking freedom: it measures the ease with which it is possible to create, develop and close a business. 
\item 9. Monetary freedom: it measures price stability in relation to a price control. 
\item 10. Labor Code\footnote{ This item has been added in 2007. Moreover,  in 2017, the Heritage Foundation made some methodological changes;  the IEF has 12
components   nowadays. The new components are “Judicial Effectiveness” belonging to the Rule of Law pillar and “Fiscal Health” as the new factor of the Government size pillar.}: it measures the ease with which workers and companies interact without restriction from the state government.
 \end{itemize}
 
 Some of these components are the results of an assembly of additional measures.
Each of these components is measured on a scale of  0 to 100. The value 100 represents the maximum freedom. The index was  obtained in
averaging these 10 components (with an equal  weight  for each of them).

Notice that more  recently,  
 Dialga and  Vallée \cite{Dialga21} 
 dealt with "methodological issues in the Index of Economic Freedom",  
indicating that  two  components,  “1. Tax  Freedom”  and 
“Government  Spending”,  which  define  the  “2. Government Size”  pillar,  are  negatively 
correlated to the other "pillars", whence making the index very unstable, - thus impairing the country ranking.

 \subsection{Economic growth}
 
Most empirical studies, e.g.,  \cite{[9],[10], [11], [12], [13]}  provide evidence that economic freedom, as measured by the Economic Freedom of the World Index, is related to economic growth, income,   standard of living, low corruption, etc.  Much evidence shows that economic freedom leads to economic growth even where countries have limited political freedom \cite{Holcombe98,DeHaan00,Doucouliagos06}. The reverse is not true.  
The case of IEF is less studied \cite{Hristova}. 
In most cases, the question turns upon the level of importance of the various independent variables.

 One of the first papers that explored the relationship between EF and growth was by  Islam  \cite{Islam96}. The first study concerning the analysis of the link between different components of EF and economic growth seems due to Ayal and Karras \cite{AyalKarras98}. However identifying which aspects of EF are more conducive to growth has proven difficult, due to multicollinearity among the index areas \cite{RodeColl}. Due to the more basic aim of our paper, we will not discuss any further regression models nor (Granger) causality in the freedom-growth relationship, here, whence reducing to an finite level tour literature review. 
 Nevertheless, for some completeness, let us point out to a few papers, either considering EF-EG from the EFW \cite{Cole03contribution,Dawson03causality} or the IEF \cite{Hristova} point of view.
   
\subsection{ 
Criticism/Limitations}
These types of indices are often criticised for their methodology. Some "economists" criticise the    economic basis on which such indices are based. They consider the measures to be too restrictive and demand that they should include a broader range of freedom concepts. Others, as John Miller  \cite{[19]}, argue that the relationship found for example between a high life  level  and such indices is the biased result of   choices made in the construction of some index. Others, like Heckelman and Stroup  \cite{[20]}, criticise the  method used in order to average components, - which they  consider to be arbitrary. See also the previous mention of Dialga and  Vallée recent finding \cite{Dialga21}.

\section{ 
Data}  \label{Data}
In order to study the spread of EF around the world, its evolution during this past decade, and subsequently  its impact on the richness of the world, it is necessary to obtain  the values of the EFW index and of the IEF  together with the gross domestic product for the studied countries

The EFW index values,  obtained from the portal $www.freetheworld.com$ \cite{[23]}, are provided for 140 countries in the 2000-2006 period, i.e.  over 7 years.
The values of the IEF  can be found on the site of the "Heritage Foundation"  \cite{[24]}.
The indices are given for 157 countries in the (12 years) period 1997-2007. The values of the Gross Domestic Product per capita (GDP)  of countries for corresponding  periods 
may be downloaded from the IMF website \cite{[25]}. All   values are annual data.  

We point out that it was alas necessary to exclude certain countries for which the data was unavailable
for various reasons. This is, for example, the case of Iraq.
Iraq’s second war has made the measurement of economic indicators quite dubious:
the values obtained for the IEF and EFW indices or for GDP could not be considered to be significant. 
 That being said, there are still 908 data points for the EFW index and 
 1784 for the IEF for the studied periods.

\subsection{ 
Statistical Characteristics of Indices Distribution  }
\begin{figure}   
    \begin{center}
\includegraphics[scale=.630] {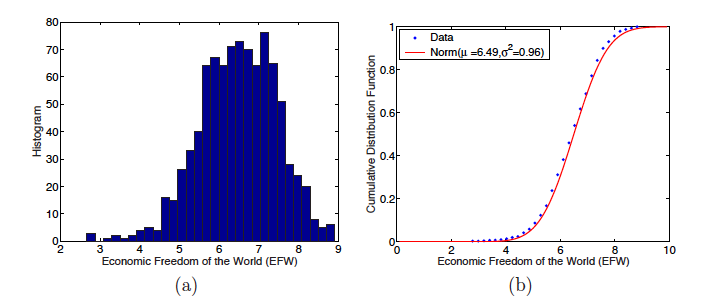}
\caption{ (a) Economic Freedom of the World (EFW) histogram for  908 data points, i.e. when available  for all  (140) 
countries and for all (7) years;  (b) cumulative probability density  for the EFW
and normal distribution fit with mean  $\mu$  = 6.49 and variance $\sigma^2$ = 0.96. }
\label{Fig2ab}
\end{center}
\end{figure}

\begin{figure}   
    \begin{center}
\includegraphics[scale=.630] {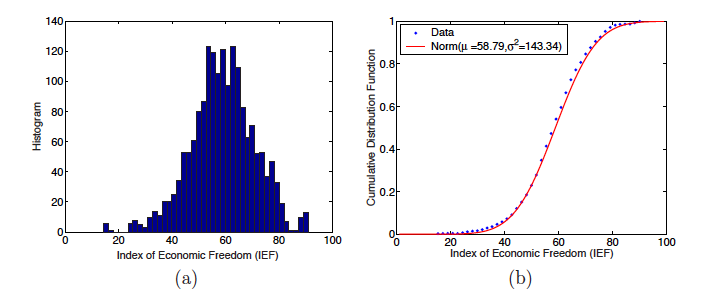}
\caption{ (a) Index of Economic Freedom (IEF) histogram for  
1784 data points;
 (b) cumulative probability density for the IEF
and normal distribution  fit with mean $\mu$ = 58.79 and variance $\sigma^2$ = 143.34. }
\label{Fig3ab}
\end{center}
\end{figure}

The first step in the study of the indices concerns the distribution of their values.
The histograms and cumulated probability densities   of the EFW and of the IEF are reproduced in Fig. \ref{Fig2ab} 
and Fig. \ref{Fig3ab}, respectively. The  main statistical characteristics
(mean, standard deviation, variance, coefficient of variation, skewness and kurtosis) of these distributions are included in  Table   \ref{TablestatEFWIEF}. 

 Fig. \ref{Fig2ab} 
and Fig. \ref{Fig3ab} 
 suggest that both indices follow a normal law slightly displaced to the right, i.e. to values greater than the median values, whence the negative skewness. This impression is reinforced by the average values of the indices: 6.49 for the EFW index and 58.79 for IEF, see  Table  \ref{TablestatEFWIEF}. 
 These two averages are greater than the corresponding median values : 5 in the case of EFW and 50 in the IEF.  The skewness is negative for both indices: -0.3567 for the EFW index and -0.2373 for the IEF, see  Table  \ref{TablestatEFWIEF},  
confirm that the probability densities are no longer  important for values above the median. These  
 features   show that the economies of the studied countries  are generally freer than constrained.

   \begin{table} \begin{center} 
\begin{tabular}[t]{|c||c|c|c|c|c|c|c|c|c|}
  \hline
  &   $\Delta T$& $N$  &   Mean    &   St.Dev.   &   Var.&CoV.& Skewn.    &   Kurt.    \\
      &     (yrs)   &    &($\mu$) &   ($\sigma$)  &     ($\sigma^2$)   &   ($\sigma/\mu$)  &  & \\\hline
EFW  &7&908   &   6.49 &   0.98 &   0.96  &    0.151 &-0.3567 & 3.3670\\
IEF  &12& 1784&   58.79  &   11.97 &   143.34 &   0.2036  &-0.2373&3.5416 \\\hline
 \end{tabular} 
   \caption{ 
    Summary of  (rounded) main statistical characteristics  of the economic freedom indicators distributions, i.e.,  the Economic Freedom of the World (EFW) index and Index of Economic Freedom (IEF),   
    according to  
    the examined time interval $\Delta T$ for the number $N$ of data points.
   }  
   \label{TablestatEFWIEF}
\end{center} \end{table}

In order to confirm that the distributions follow a normal law, a Kolmogorov-Smirnov (KS) test is performed. The results of the tests are shown in  Table \ref{TableKS}. 
The  KS distances, DKS = 0.0399 for EFW and 0.0310 for IEF,   are lower than the "critical values"  of the normal distribution,  0.0449 for EFW and 0.0321 for the IEF.
 In addition, $p-$values, 0.1088 for the EFW index and 0.0633 for the IEF,  are above the   5\%
significance level; thus the KS tests are considered to lead to statistically significant features.  
It is therefore possible to conclude that the EFW index and IEF  values follow a  normal law  with $\mu$ = 6.49 and 58.79 and variance $\sigma^2$ = 0.96 and 143.34 respectively, i.e. the standard distribution (SD) is equal to 0.98 and 11.98, respectively.

  \begin{table} \begin{center} 
\begin{tabular}[t]{|c|cc|cc|}
  \hline
Kolmogorov-Smirnov (KS) test&$EFW$& $IEF$\\ \hline
$p-$value  &0.1088&  0.0633 \\
Gaussian Distribution Critical Value& 0.0449&  0.0321\\
Significance Level &0.05&  0.05\\
Number of data points  &908 &1784\\
\hline
$DKS$&0.0399&0.0310\\ 
\hline
 \end{tabular}  
   \caption{ 
   Kolmogorov-Smirnov (KS) test for the adjustment of data from EFW and IEF to a normal distribution. 
The distances of KS (DKS) and  the $p-values$ indicate that KS tests are statistically significant. It is therefore allowed to conclude that the EFW and IEF  values follow a normal law,  with $\mu$= 6.49 and 58.79 and variance  $\sigma^2$  =  0.96 and 143.34, respectively.
  } 
   \label{TableKS}
\end{center} \end{table}

\subsection{
EFW Index  in Year 2006 }

For example, consider a specific year, 2006.  
Table \ref{TableEFW20r} 
shows the EFW index values for the 20 freest countries   for the year 2006. Hong Kong, Singapore and New Zealand
occupy the first 3 places. The rest of the top 20 is made up of the great Anglo-Saxon countries (USA, Canada, Australia) and European countries (Switzerland,
United Kingdom, Ireland, Estonia, Iceland, Denmark, Finland, Austria, Netherlands, Germany, Slovakia). It should be noted that there is one South American country,
Chile (in 6th position) and one country from the Arabian Peninsula, Kuwait (in 19-th position).

In constrast, Table \ref{TableEFW121+r} 
 shows the EFW index for the 21 least free countries   in 2006.  
It is remarkable that the least free countries are mainly regrouped in Africa:  16 out of the 21 last countries.

  \begin{table} \begin{center}   
\begin{tabular}[t]{|ccc|ccc|}
  \hline
  \multicolumn{6}{|c|}{ $2006$ $EFW$ $ ranking$}  \\ \hline
$rank$&Country&$2006$	& $rank$ &Country&$2006$ \\
\hline
1& Hong-Kong& 8.94 	&11& Estonia& 7.89\\
2 &Singapore& 8.57 	&12& Iceland& 7.8\\
3 &New Zealand& 8.28 	&13& Denmark& 7.78\\
4 &Switzerland& 8.20 	&14& Finland& 7.69\\
5 &United Kingdom& 8.07 	&15& Austria& 7.66\\
6 &Chile& 8.06 	&16& Netherlands& 7.65\\
7 &Canada& 8.05 	&17& Germany& 7.64\\
8 &Australia& 8.04 	&18& Taiwan& 7.63\\
8 &United States& 8.04 &19& Kuwait& 7.62\\
10 &Ireland& 7.92 	&20& Slovak Rep.& 7.61\\
\hline
 \end{tabular}   
   \caption{ 
    2006 Economic Freedom of the World (EFW) Index values for the 20 freest countries. 
 }  
   \label{TableEFW20r}
\end{center} \end{table}

  \begin{table} \begin{center} 
\begin{tabular}[t]{|ccc|ccc|}
  \hline
  \multicolumn{6}{|c|}{ $2006$ $EFW$ $ ranking$}  \\ \hline
$rank$&Country&$2006$	& $rank$ &Country&$2006$ \\
\hline
121& Ethiopia &5.64 &131& Burundi& 5.23\\
121& Ukraine &5.64 &131& Rwanda& 5.23\\
123& Burkina Faso &5.63 &133& Chad& 5.12\\
124& Algeria &5.57 &134 &Central Africa Rep.& 5.01\\
125& Syria &5.54 &134 &Guinea-Bissau &5.01\\
126& Malawi &5.42 &136 &Venezuela& 4.76\\
127& Gabon &5.37 &137 &Niger &4.67\\
128& Nepal &5.35 &138 &Congo, Rep. of& 4.64\\
129& Togo &5.33 &139 &Myanmar& 4.19\\
130& Congo, Dem. Rep. &5.25 &140& Angola& 4.10\\
& & &141& Zimbabwe& 2.67\\
\hline
 \end{tabular}  
   \caption{ 
   2006 Economic Freedom of the World (EFW) Index values for the 21 least free countries. Unlike the 20 freest countries  on the planet, the 21 least  free countries are almost all   in Africa (16 of the 21).  }
   \label{TableEFW121+r}
\end{center} \end{table}

\subsection{
IEF in Year 2006 }

Similarly, Table  \ref{TableIEF20r} and Table   \ref{TableIEF138+r}
  list the IEF values for the 20 freest countries and the 20 least free countries, respectively.
The former British colonies still dominate the ranking.  Hong Kong and Singapore occupy the top 2 places in the ranking. The big  Anglo-Saxon (United States, United Kingdom, Australia and Canada) countries are also in the top 20. Among all the regions of the world, Europe has the largest number of  countries in the top 20 (9 of the 20 countries are European).

As in the case of the EFW index, a large majority of the "less free" countries are in Africa (10 out of 20 countries). The (last) Communist Countries (North Korea and Cuba) are appearing in the 2 last places of the ranking.

  \begin{table} \begin{center}
\begin{tabular}[t]{|ccc|ccc|}
  \hline
  \multicolumn{6}{|c|}{ $2006$ $IEF$ $ ranking$}  \\ \hline
$rank$&Country&$2006$	& $rank$ &Country&$2006$ \\
\hline
1& Hong-Kong& 88.6& 11 &Iceland &75.8\\
2 &Singapore& 88.0 &12& Denmark& 75.4\\
3 &Ireland& 82.2 &12 &Netherlands, The& 75.4\\
4 &New Zealand& 82.0 &14& Luxembourg& 75.3\\
5 &United States& 81.2 &15 &Estonia& 74.9\\
6 &United Kingdom& 80.4 &16 &Japan& 73.3\\
7 &Australia& 79.9 &17 &Finland& 72.9\\
8 &Switzerland& 78.9 &18& Bahamas, The& 72.3\\
9 &Chile& 78.0 &19& Barbados& 71.9\\
10 &Canada& 77.4 &20 &Cyprus& 71.8\\
\hline
 \end{tabular}  
   \caption{
   2006  Index of Economic Freedom (IEF) values for the 20 freest countries.   
 }  
   \label{TableIEF20r}
\end{center} \end{table}

  \begin{table} \begin{center} 
\begin{tabular}[t]{|ccc|ccc|}
  \hline
  \multicolumn{6}{|c|}{ $2006$ $IEF$ $ ranking$}  \\ \hline
$rank$&Country&$2006$	& $rank$ &Country&$2006$ \\
\hline
138& Chad &50.0 &148 &Iran& 45.0\\
139& Haiti &49.2 &149 &Venezuela& 44.6\\
140& Nigeria& 48.7& 150& Turkmenistan& 43.8\\
140& Burundi &48.7 &150 &Congo. Rep. of &43.8\\
140& Uzbekistan& 48.7& 152& Angola& 43.5\\
143& Laos &47.5 &153 &Burma &40.0\\
143& Belarus& 47.5 &154 &Zimbabwe &33.5\\
145& Togo &47.3 &155 &Libya &33.2\\
146& Guinea-Bissau &46.5 &156 &Cuba& 29.3\\
147& Sierra Leone &45.2 &157 &Korea. North& 4.0\\
\hline
 \end{tabular}  
   \caption{ 
   2006  Index of Economic Freedom (IEF) values for the 20 least free countries. 
    }  
   \label{TableIEF138+r}
\end{center} \end{table}

\subsection{
Regional Evolution of Economic Freedom  }

In order to study the geographical distribution of economic freedom,
 it is possible to calculate an "average freedom value" for  the six major continents (Africa,
North America, South America, Asia, Europe and West Africa) \footnote{ The distribution of countries by continent is carried out by following the geographical scheme of  the United Nations Statistics Division  \cite{[26]}.  This partition has been chosen because it has been  developed with the aim of conducting statistical studies relevant to the various regions.}. 
However, the calculation of such an  average selected is not a simple arithmetic mean. It does not make sense to give a similar weight to the United States and  e.g., to Ecuador, to  China or to Vietnam. Instead, we consider that the weight should depend on the country contribution to the world economy, - for example through the GDP.
Thereafter, the weight is given by
\begin{equation}
w_i = \frac{GDP_i }{\sum_{j=1}^N GDP_j}
\end{equation}
where  $w_i$ represents the weight of the country $i$ and $GDP_j$, the internal product country $j$.

\begin{figure}   
    \begin{center}
\includegraphics[scale=.630] {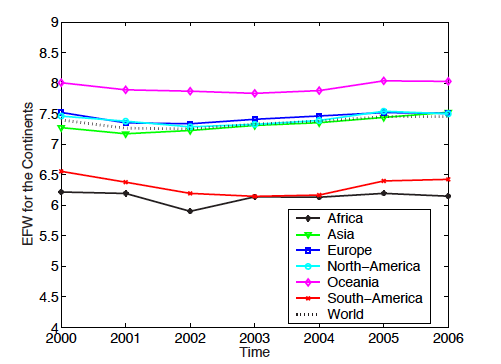}
\caption{  Yearly evolution  of the Economic Freedom of the World (EFW)  Index for the six continents (Africa, Asia,
Europe, North America, West Africa and South America).  The
index calculation for a region results from a weighted averaging of the indices of the
countries belonging to the specific region. The weight of  a country is the ratio of the
GDP of the  country to  the GDP of the world economy.  }
\label{Fig4}
\end{center}
\end{figure}

\begin{figure}   
    \begin{center}
\includegraphics[scale=.630] {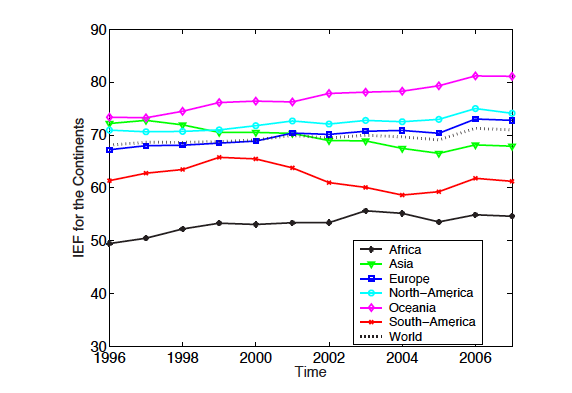}
\caption{ Yearly evolution  of the Index of Economic Freedom (IEF) for the six continents (Africa, Asia,
Europe, North America, West Africa and South America).  The
index calculation for a region results from a weighted averaging of the indices of the
countries belonging  to the specific region. The weight of  a country is in the ratio of the
GDP of the country to  the GDP of the world economy. }
\label{Fig5}
\end{center}
\end{figure}

The evolutions of the EF  for  the 6 continents, obtained by this method  are reproduced in Fig. \ref{Fig4} 
 for the EFW index, and  in Fig. \ref{Fig5} 
 for the IEF.

For the EFW index, Fig. \ref{Fig4} 
shows that   Oceania is the  the freest of the six regions, with an index value $\simeq 8$ , relatively stable of the 7 years.  Europe, North America and Asia are {\it ex aequo} with a value $\simeq 7.5$, which represents the world average value. Africa is the  less free region and   South America does not make much better.

For the IEF, Fig. \ref{Fig5} 
also shows that   Oceania  is the   freest region with an ever increasing value. It goes   from
73.36 in 1996  to 81 in 2007. Europe and North America follow the same evolution and have almost identical values. Asia regresses in terms of "economic freedom", 
even though there is a slight  improvement  the last two years.
It goes from 72.2  to 67.9 with a minimum value equal to 66.4 in 2005. Africa is again the least free region of the world, but progresses over the
12 years period. Overall, the world average freedom is rising from 68 in 1996 to 71 in 2007.

The   "rate changes" appear  to be different from one index to the other; this is due to  the periods of study. Indeed, if the study period is restricted to 2000-2006 for the IEF, the results so obtained for both indices are almost identical. The slight differences are explained by the fact that the IEF is "more conservative" than the EFW; the IEF leads to values lower than EFW for a country.  Further discussion on this topic is  postponed to section 
\ref{comparisonofindices}.
 
\subsection{
Exponential  versus power law behaviour}

In this section, countries are ranked according to the value of the indices in a conventional order:
a low ranking indicates that the country belongs to the group of the freest countries in the world.
Conversely, a "high" rank means that the country has an index value, whence a low EF as  compared to others.

The goal here is to determine, the so called ''rank-size'' law, once the  countries are ranked,  in particular whether the indices follow
an exponential   or a power law\footnote{These are the two most simple analytical functions carried over from statistical physics to econophysics; whence their  mathematical origin is well known and not further  discussed.}, i.e.,
\begin{equation}
INDEX \sim e^{\lambda r}  
\end{equation}
or 
\begin{equation}
INDEX \sim r^\nu  
\end{equation}
where $r$ is the rank of the country;  $\lambda$  and $\nu$  are  characteristic exponents. The latter equation corresponds to the (so called Zipf) rank-size law \cite{r15}, if $\nu=-1$.

\begin{figure}   
    \begin{center}
\includegraphics[scale=.730] {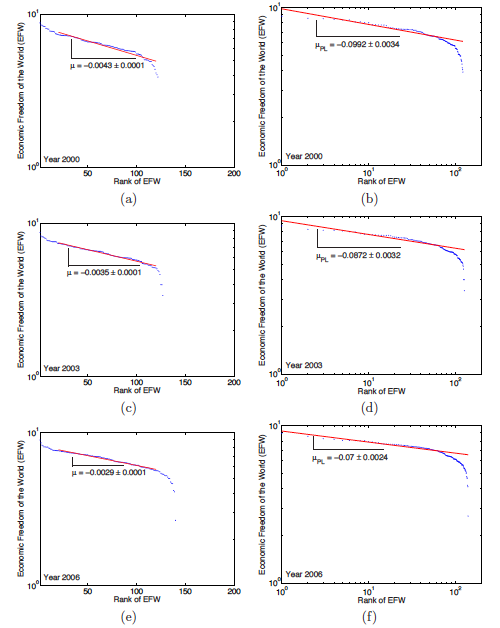}
\caption{  Examples of semi-log [(a), (c), (e)] and log-log [(b),  (d), (f)] plots of the rank-size  relation
between the Economic Freedom of the World (EFW)  index and the country rank for the years 2000, 2003 and 2006,
respectively:  the semi-log plots show that the relationship is exponential
for countries of high rank ($\ge  20$);  the log-log plots point to a behaviour close to a power law. }
\label{Fig6}
\end{center}
\end{figure}

 Fig. \ref{Fig6} 
  (a), (c) and (e) show that the EFW has an  exponential behaviour  for countries with a  rank higher than
20. The value  of the exponent decreases a little bit more each year and ends up to stabilise  at $\simeq$ -0.0049 in 2005 and 2006 (see  Table \ref{TablelambdaEFWr}). The low error bars (less than  0.0001) and the high value regression coefficient  (the regression coefficient is greater than 93\%) confirm that the data perfectly follow the exponential law.

  Fig. \ref{Fig6} 
  (b), (d) and (f) show the power-law behaviour of the EFW.  Table  8 reports the values of the exponent of the power law  
 for the 6 studied years. It does not vary much between 2001  and  2004; it falls to -0.0743 in 2005 and -0.007 in 2006. Here again, the effectiveness of the
regressions is  high, between $\sim $ 89 and 93\%. This indicates that the data follows a  power law.

\begin{figure}   
    \begin{center}
\includegraphics[scale=.730] {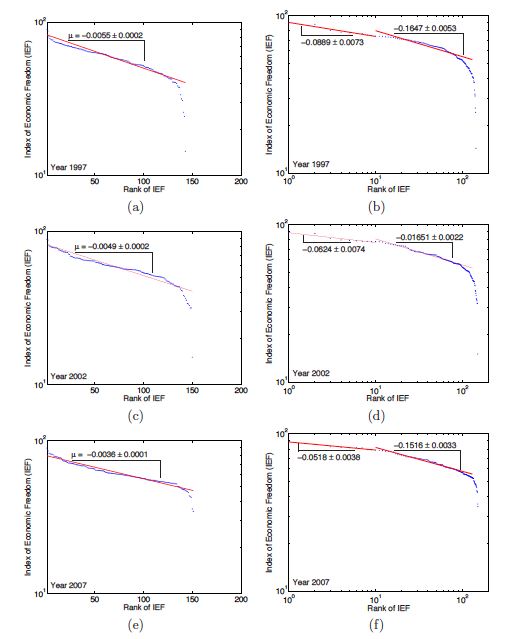}
\caption{ Examples of semi-log  [(a), (c), (e)] and log-log [(b),  (d), (f)] plots of the rank-size  relation
between the  Index of Economic Freedom (IEF)  and the country rank for the years 1997, 2002 and 2007,
respectively; the semi-log plots show that the IEF  ranking has an exponential behaviour;  the log-log plots point to 
the existence of a transitional point between two different power laws, i.e., near rank 10.  }
\label{Fig7}
\end{center}
\end{figure}

For the IEF, the semi-log graphs, see Fig. \ref{Fig7} 
(a), (c) and (e), point to an exponential behavior according to the rank of countries.
The exponent  decreases every year, going down from -0.006 in 1996 to -0.0036 in 2007 
(see Table \ref{TablelambdaIEFr}). The regression coefficient shows that the   exponential law  has been  "perfectly"   followed since 2003, a year  for which the efficiency of the regression exceeds 90\%.

 \begin{table} \begin{center}    
\begin{tabular}[t]{|c||c|c|c|c|}
  \hline
 &  \multicolumn{4}{|c|}{ $EFW\sim e^{\lambda r}$}  \\ \hline
year&$\lambda$&$\Delta \lambda$&$\Delta \lambda/\lambda$&$ R^2$\\
\hline
2000& -0.0043& 0.0001& 0.0272 &0.9316\\
2001& -0.0039& 0.0001&  0.0257 &0.9388\\
2002& -0.0037& 0.0001&  0.0208 &0.9591\\
2003& -0.0035& 0.0001&  0.0129 &0.9839\\
2004& -0.0035& 0.0001&  0.0068 &0.9954\\
2005& -0.0029& 0.0001&  0.0107 &0.9889\\
2006& -0.0029& 0.0001&  0.0113 &0.9876\\
\hline
 \end{tabular} 
   \caption{
   Yearly evolution of the $\lambda$ exponent in the assumed empirical exponential  law  between the EFW  index
   and the rank ($r$),  the standard error
($\Delta \lambda$), its relative value ($\Delta \lambda/\lambda$),  and the efficiency ($R^2$) of the regression.  The low error bar values (less than 0.0001) and the effectiveness of the regressions  ($\ge  93\%$) confirm that the data are perfectly following the exponential law.
   }  
   \label{TablelambdaEFWr}
\end{center} \end{table}

Unlike the EFW index, for which the data follow a power law  for all ranks,  Fig. \ref{Fig7} 
(b), (d) and (f) show a transition point between 2 different power law for the IEF, near    rank  10 do.   The exponent of the law for countries with a rank below 10 "increases" over the years, from -0.0931 in 1996 to -0.0518 in 2007.  
The exponent  for countries  with rank higher than 10 remains relatively stable $\simeq$ -0.016 over the 12 years  here studied (see  Table \ref {TablenuIEFr}). 

It should be noted that countries with low EF (those which have a very high rank) do not follow neither a  power law nor an exponential law; this feature holds  for  both indices. The difficulty of performing economic measures for these countries can explain  that the index values are fraught with errors that are
 not possible to compensate. These countries are often those with very little developed economy,  weakly connected to their outside world, apparently subject to the will of a dictator.
 
 \subsection{
 Comparison of both indices} \label{comparisonofindices}
The purpose of this section is to compare the indices, whence it is necessary to restrict the  observation "period" at the largest  but common year interval.  We should also take into account the countries common to both sets. That leaves 138 countries to be examined  over a period extending from 2000 to 2006, i.e. 2 sets of 862 data points.

To have a meaningful comparison, it is best to "normalise" the index values in a  observation interval; here we choose the interval  to be [0,1]. To do so,
it is sufficient to divide the values of the EFW index by 10 and those of the IEF  by 100.

\begin{figure}   
    \begin{center}
\includegraphics[scale=.830] {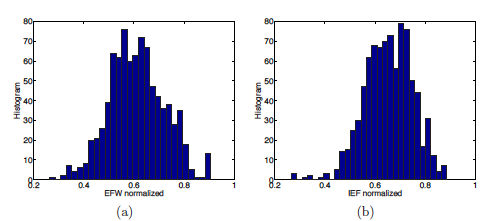}
\caption{ Histogram of (a) Economic Freedom of the World (EFW) and (b)  Index of Economic Freedom (IEF)  values for the 862 data points, common to both indices,  normalised over [0,1]. }
\label{Fig8ab}
\end{center}
\end{figure}

The distributions of the 862  data points   are reproduced in  Fig. \ref{Fig8ab} 
for both indices. The average of the EFW values  is  
$\simeq  0.6542$, while the average for  the IEF is  slightly lower   
$\simeq  0.6118$. This shows that the EFW gives, on average, an index value   slightly greater than that given by the IEF for the same country (see  Table \ref{TablestatEFWIEFn}). 

\begin{figure}   
    \begin{center}
\includegraphics[scale=.630] {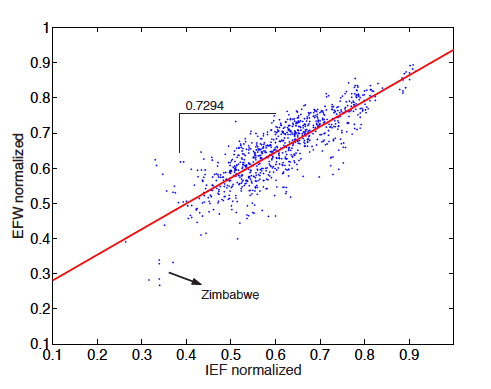}
\caption{ Scatter plot of the relationship between the Economic Freedom of the World (EFW)  index and the  Index of Economic Freedom (IEF)  normalised values. 
The regression slope  points to a linear relationship $\sim$ 0.7294. This value,  statistically significant,  lower than
 1, confirms that the IEF is   "more conservative" than the EFW index. The  worst EFW country  (Zimbabwe) position is emphasised for  framing the data. }
\label{Fig9}
\end{center}
\end{figure}

In order to confirm that the IEF is more conservative than the EFW index, it is interesting to represent the EFW values according to the IEF values.
This is done in   Fig. \ref{Fig9}. 
By calculating the linear regression coefficient, the  slope is found to be 0.7294.
 This value is markedly less than 1, whence  confirming that the EFW gives   index values greater than  the IEF for a given country.

 \begin{table} \begin{center} 
\begin{tabular}[t]{|c||c|c|c|c|}
  \hline
 &  \multicolumn{4}{|c|}{ $EFW\sim r^{\nu }$}  \\ \hline
year&$\nu$&$\Delta \nu$&$\Delta \nu/\nu$&$ R^2$\\
\hline
2000& -0.0992& 0.0034& 0.0343 &0.9161\\
2001& -0.0907& 0.0029& 0.0314 &0.9285\\
2002& -0.0890& 0.0029& 0.0328 &0.9226\\
2003& -0.0872& 0.0032& 0.0369 &0.9038\\
2004& -0.0857& 0.0034& 0.0393 &0.8924\\
2005& -0.0743& 0.0023& 0.0306 &0.9319\\
2006& -0.0700& 0.0024& 0.0344 &0.9154\\
\hline
 \end{tabular}
   \caption{
   Yearly evolution of the $\nu$ exponent in the  empirical power law  between the EFW     and the rank ($r$),  the standard error
($\Delta \nu$), its relative value ($\Delta \nu/\nu$), and the efficiency ($R^2$) of the regression.  The low error bar values  ($\Delta \nu/\nu$ $\simeq  3 \%$)   and the effectiveness of the regressions
  confirm that the data is  well following  a power law.
   } 
   \label{TablenuEFW}
\end{center} \end{table}

\section{
Relationship between economic freedom and wealth of countries} \label{relationships}
As recalled here above, many studies show a strong relationship between economic freedom and the wealth of a country, i.e. between EF 
and the country gross domestic product (GDP). In this section, the goal is to evidence  this relationship.  

\begin{figure}   
    \begin{center}
\includegraphics[scale=.630] {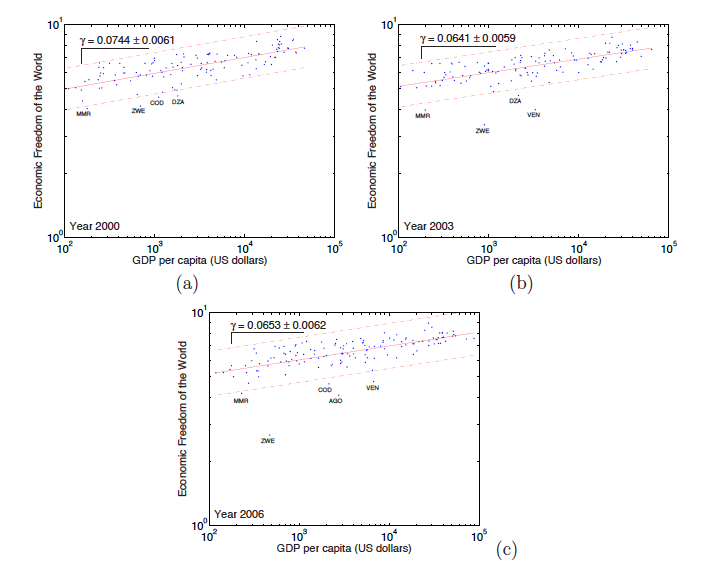}
\caption{ Examples of  Log-Log plot of the Economic Freedom of the World (EFW)  Index  with respect to
  country's gross domestic product (GDP) for the years (a) 2000, (b) 2003 and (c) 2006.
 This relationship is characterised by a power law, with an exponent $\gamma \simeq 0.674$. The
 doted lines encompass the region for which the data is within twice the standard deviation away from the trend. }
\label{Fig10abc}
\end{center}
\end{figure}

\begin{figure}   
    \begin{center}
\includegraphics[scale=.630] {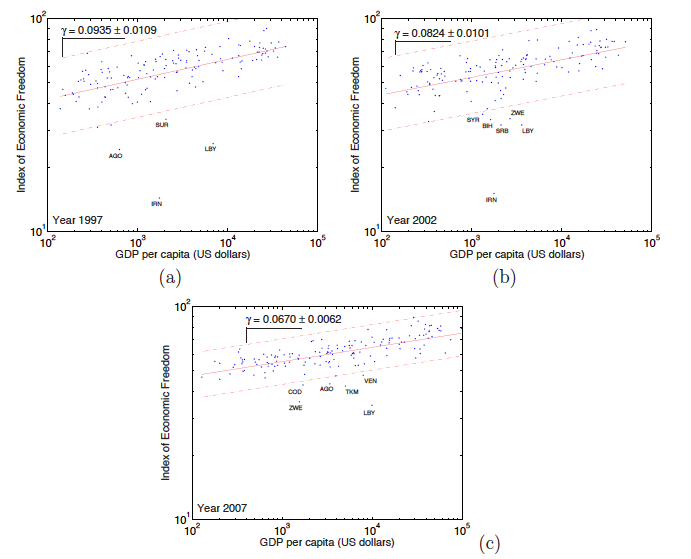}
\caption{ Examples of Log-Log plots of the  Index of Economic Freedom (IEF)  relationship to the country's gross domestic product (GDP) for the years (a) 1997, (b) 2002 and (c) 2007.
 This relationship is characterised by an evolutive  power law. The  doted lines limit the region for which the data are located within a maximum distance equal to twice the standard deviation; the  few outliers  have been removed for calculating the power law exponent  $\gamma$. }
\label{Fig11abc}
\end{center}
\end{figure}

A graphic representation of EF according to the GDP,  on Fig. \ref{Fig10abc}  and Fig. \ref{Fig11abc},  
shows  that the    relationship translates into a power law, i.e., thereby defining the exponent $\gamma$, 
  \begin{equation}
  INDEX \simeq GDP^\gamma \; .
\end{equation}
A positive exponent ($\gamma  > 0$)   indicates a "positive relationship" between EF and  the GDP. 
  This would mean that the freest countries are the richest ones.
A negative exponent  indicates a negative correlation:   the freest countries would be the less rich ones.

The existence of this law is very important from an economic point of view. Indeed, it
   allows us to know the wealth  which a country  should have  as a function of    its level of economic freedom. By estimating the influence that  
a government decision will have on the economic freedom index of that country, it is  possible to measure directly the impact of a government policy
on the economy of the country.
Moreover, the existence of this (simple)  law will enable countries to be classified according to their   position on   the power law.  Countries that are located  above the law are countries that have a lower gross domestic product than   that they should have for their level of economic freedom. These  countries can be said to be    ‘underperforming’. 

On the other hand, the countries that are located below the law are countries that have a  gross domestic product greater than that which it should have. These countries are ‘over-performing’.

 \begin{table} \begin{center} 
\begin{tabular}[t]{|c||c|c|c|c|}
  \hline
 &  \multicolumn{4}{|c|}{ $IEF\sim e^{\lambda r}$}  \\ \hline
year&$\lambda$&$\Delta \lambda$&$\Delta \lambda/\lambda$&$ R^2$\\
\hline
1996 &-0.0060& 0.0003& 0.0422 &0.8087\\
1997 &-0.0055& 0.0002& 0.0405 &0.8124\\
1998 &-0.0057& 0.0002& 0.0385 &0.8211\\
1999 &-0.0054& 0.0002& 0.0416 &0.7919\\
2000 &-0.0051& 0.0002& 0.0382 &0.8185\\
2001 &-0.0050& 0.0002& 0.0345 &0.8508\\
2002 &-0.0049& 0.0002& 0.0381 &0.8235\\
2003 &-0.0044& 0.0001& 0.0212 &0.9374\\
2004 &-0.0043& 0.0001& 0.0241 &0.9215\\
2005 &-0.0041& 0.0001& 0.0246 &0.9180\\
2006 &-0.0037& 0.0001& 0.0238 &0.9223\\
2007 &-0.0036& 0.0001& 0.0227 &0.9285\\
\hline
 \end{tabular}
   \caption{ 
   Yearly evolution of the $\lambda$ exponent in the  empirical exponential law  between the IEF
   and the rank ($r$),  the standard error
($\Delta \lambda$), its relative error ($\Delta \lambda/\lambda$), and the efficiency ($R^2$) of the regression.  The low error bar values  ($\Delta \lambda/\lambda$ $\simeq  2$ to $4 \%$)   and the effectiveness of the regressions
  confirm that the data are  well following  a power law.
   }  
   \label{TablelambdaIEFr}
\end{center} \end{table}

On Table \ref{TablelambdaIEFr}, we report the exponential law parameter ($\lambda$) between the IEF and the rank ($r$) of the IEF, the Standard 
Error ($\Delta \lambda$)  and its Relative Error ($\Delta \lambda/\lambda$), together with the efficiency of the regression ($R^2$). The $\lambda$ value decreases each year  (in absolute value); it increases from -0.006 in 1996 to -0.0036 in 2007. 
The efficiency of the regression shows that the data follow an exponential law, rather  perfectly since 2003, when the efficiency of the regression exceeds 90\%.
 
  \begin{table} \begin{center} 
\begin{tabular}[t]{|c||c|c|c|c||c|c|c|c|}
  \hline
 &  \multicolumn{8}{|c|}{ $IEF\sim r^{\nu}$}  \\ \hline
  &  \multicolumn{4}{|c||}{ $r\le 10$}&  \multicolumn{4}{|c|}{ $r \in [10-100]$}    \\ \hline
year&$\nu$&$\Delta \nu$&$\Delta \nu/\nu$&$ R^2$&$\nu$&$\Delta \nu$&$\Delta \nu/\nu$&$ R^2$\\
 & & &$(\%)$&$(\%)$ & &&$(\%)$&$(\%)$\\
\hline
1996& -0.0931 &0.0071 & 7.60 & 95.58 &-0.1820 &0.0056 & 3.09 & 92.16\\
1997& -0.0889 &0.0073 & 8.26 & 94.82 &-0.1647 &0.0053 & 3.25 & 91.43\\
1998& -0.0808 &0.0099 & 12.28 & 89.24 &-0.1505 &0.0044 & 2.92 & 92.96\\
1999& -0.0797 &0.0079 & 9.90 & 92.73 &-0.1477 &0.0029 & 1.95 & 96.73\\
2000& -0.0807 &0.0089 & 10.97 & 91.21 &-0.1504 &0.0030 & 1.98 & 96.63\\
2001& -0.0723 &0.0058 & 8.02 & 95.11 &-0.1634 &0.0042 & 2.54 & 94.57\\
2002& -0.0624 &0.0074 & 11.80 & 89.97 &-0.1651 &0.0022 & 1.36 & 98.38\\
2003& -0.0686 &0.0070 & 10.17 & 92.35 &-0.1704 &0.0031 & 1.81 & 97.18\\
2004& -0.0690 &0.0092 & 13.35 & 87.52 &-0.1690 &0.0024 &1.45 & 98.16\\
2005& -0.0717 &0.0095 & 13.26 & 87.67 &-0.1678 &0.0024 & 1.40 & 98.28\\
2006& -0.0564 &0.0047 & 8.29 & 94.78 &-0.1522 &0.0022 & 1.45 & 98.17\\
2007& -0.0518 &0.0038 & 7.33 & 95.88 &-0.1516 &0.0022 &1.47 & 98.11\\
\hline
 \end{tabular}  
   \caption{ 
   Yearly evolution of the Zipf   law exponent ($\nu$) between the IEF and the rank ($r$) of
IEF, the Standard Error ($\Delta \nu$), the Relative  Standard   Error $(\Delta \nu/\nu$),   and the Regression  Coefficient ($R^2$). While the exponent  for countries of rank  below 10  decreases over the years the exponent  for countries of  rank higher than 10 remains relatively stable, near the value -0.016 over the 12 years of the study.  }  
   \label{TablenuIEFr}
\end{center} \end{table}

 On Table \ref{TablenuIEFr}, 
we report the (Zipf)   rank-size law exponent ($\nu$) between the IEF and the rank ($r$) of IEF, the Standard Error ($\Delta \nu$), the Relative  Standard   Error ($\Delta \nu/\nu$),  and the yearly regression  coefficients ($R^2$), for the observed different regimes. While the exponent  for countries of rank  below 10  decreases over the years the exponent  for countries of  rank higher than 10 remains relatively stable, near the value -0.016 over the 12 years of the study. 

 \begin{table} \begin{center} 
\begin{tabular}[t]{|c||c|c|c|c|c|c|c|c|c|c|c|c|}
  \hline
Variable &$N_c$   &   $\Delta T$&$N$&     Mean    &   StDev   &   CoV    \\
  & &     (years)   &      &($\mu$) &   ($\sigma$)   &    ($\sigma/\mu$)      \\\hline
EFW &  138&7&862 &     0.6542 &   0.0948    &    0.1449  \\
IEF & 138  &7& 862&        0.6118  &   0.1094   &   0.1788  \\\hline
 \end{tabular} 
   \caption{ 
   Summary of  (rounded) main statistical characteristics  for  the so called  ''normalized'' EFW and IEF distributions of the economic freedom indicators,   according to the number of countries $N_c$, the examined time interval $\Delta T$,  whence the number $N$ of data points.
   }  
  \label{TablestatEFWIEFn} 
\end{center} \end{table}

On Table \ref{TablestatEFWIEFn}, 
 we report the main characteristics (average and standard deviation) of the normalised   EFW and IEF  data for the 138 countries out of the 7 years (i.e. 862 data points). The EFW mean   is  slightly higher than that for  the EFW data. The coefficient of variation ($\sigma/\mu$) shows a weak dispersion in both cases.

  \begin{table} \begin{center}   
\begin{tabular}[t]{|c|c|}
  \hline
    \multicolumn{2}{|c|}{$EFW$}  \\ \hline 
year& countries\\ \hline
2000 &DZA-COD-MMR-ZWE\\
2001 &DZA-ZWE\\
2002 &DZA-COD-MMR-VEN-ZWE\\
2003 &DZA-MMR-VEN-ZWE\\
2004 &DZA-COD-VEN-ZWE\\
2005 &DZA-COD-VEN-ZWE\\
2006 &AGO-COD-MMR-VEN-ZWE\\
\hline
 \end{tabular}  
   \caption{ 
   List of countries for which the EFW Index does not comply with the power law, i.e., are  located outside the area   limited by twice the standard deviation  from the power  law. } 
   \label{Table12EFW}
\end{center} \end{table}

 On Table \ref{Table12EFW}, we list  countries\footnote{The ISO 3166-1 code is used to facilitate the presentation of data. }  fo which the EFW Index does not comply with the power law, i.e., the data points are  located outside the area   limited by twice the standard deviation  from the power  law.

  Fig. \ref{Fig10abc}  and Fig. \ref{Fig11abc} clearly show that all countries, with a few exceptions   obey the power law. The variation coefficient ($\sigma/\mu$) shows a weak dispersion on both cases, because the countries are almost all in an interval corresponding to twice the standard deviation.
For the EFW, the countries that pose a problem are  Algeria, the Republic of Congo, Burma and Zimbabwe, but also Venezuela since 2002. 
As regards the IEF, the problematic countries are  more numerous:  among these are Angola, Bosnia, Iran, Laos, Libya and Zimbabwe. Venezuela is only an IEF problem since 2004. 
The lists of such countries are included in 
Table \ref{Table12EFW} and Table  \ref{Table13IEF}    for each year of interest.  
In   Table \ref{Table12EFW}, we report the list of countries $i$ for which  the EFW Index values do not comply with the power law.
 In Table \ref{Table13IEF}, 
we report the list of countries $i$ for which  the IEF Index does not comply with the power law.  

The exponent $\gamma$  values for the period 2000 to 2006 relationship between  EFW  and GDP are  reported in   Table \ref{TablegammaEFWGDP},  
 while the $\gamma$ values  for the IEF for the 1996 to 2007 period  are shown in  Table \ref{TablegammaIEFGDP}. 
In the case of the EFW (see   Table \ref{TablegammaEFWGDP}),  
 the exponent of the law in question remains stable on the 7 years with an average value $\simeq$  0.0674.  
Notice that the  regressions coefficients for the EFW-GDP relation are not as high as in the case of the exponential and power (rank-size)  laws. 
For the IEF, there are 3 periods on the 12 years during which the exponent  holds different behaviours. For the  1996  to 2000 years,
the exponent has an average value equal to  0.0948, - which remains stable around this value over these 5 years. The second phase, which extends over the years 2001 to 2005, is a transition  period during which the value of the exponent falls down. It ends up to some stabilisation around 0.0666 during the third period  (2006-2007). The efficiency of the regressions is not very good,  except for the third period during which  $R^2$ is  approaching 50\%. Therefore, it may be  conjectured  that  the IEF corrections, added in 2006, are bearing fruit.

  \begin{table} \begin{center} 
\begin{tabular}[t]{|c|c|}
  \hline
    \multicolumn{2}{|c|}{$IEF$}  \\ \hline 
year& countries\\ \hline
1996 &AGO-AZE-IRN-LBY\\
1997 &AGO-IRN-LBY-SUR\\
1998 &AGO-BIH-IRN-LOA-LBY-UZB\\
1999 &AGO-BIH-COG-IRN-LAO-LBY-UZB\\
2000 &AGO-COG-IRN-LOA-LBY\\
2001 &BLR-BIH-LOA-LBY\\
2002 &BIH-IRN-LBY-SRB-SYR-ZWE\\
2003 &BLR-BIH-LBY–SYR-ZWE\\
2004 &BLR-LBY-SYR-VEN-ZWE\\
2005 &LBY-VEN-ZWE\\
2006 &AGO-COD-LBY-TKM-VEN-ZWE\\
2007 &AGO-COD-LBY-TKM-VEN-ZWE\\
\hline
 \end{tabular} 
   \caption{    
   List of countries fo which the IEF does not comply with the power law, i.e., are  located outside the area   limited by twice the standard deviation  from the power  law. }
   \label{Table13IEF}
\end{center}
 \end{table}

  \begin{table} \begin{center} 
\begin{tabular}[t]{|c||c|c|c|c|}
  \hline
 &  \multicolumn{4}{|c|}{ $EFW\sim GDP^{\gamma}$}  \\ \hline
year&$\gamma$&$\Delta \gamma$&$\Delta \gamma/\gamma$&$ R^2$\\
\hline
2000& 0.0744 &0.0061 &0.0824 &0.5490\\
2001& 0.0669 &0.0061 &0.0917 &0.4959\\
2002& 0.0636 &0.0062 &0.0978 &0.4636\\
2003& 0.0641 &0.0059 &0.0922 &0.4847\\
2004& 0.0705 &0.0057 &0.0814 &0.5410\\
2005& 0.0667 &0.0062 &0.0934 &0.4540\\
2006& 0.0653 &0.0062 &0.0952 &0.4443\\
\hline
 \end{tabular}  
   \caption{ 
   Yearly evolution of the  
   power law exponent ($\gamma$) between the EFW and GDP, the standard error
($\Delta \gamma$), the relative error bar  ($\Delta \gamma/\gamma$)and the efficiency ($R^2$) of the regression. The power law exponent
remains rather stable over the 7 years with an average value  $\simeq$ 0.0674 ($\pm 0.004$).    } 
   \label{TablegammaEFWGDP}
\end{center} \end{table}

   \begin{table} \begin{center}  
\begin{tabular}[t]{|c||c|c|c|c|}
  \hline
 &  \multicolumn{4}{|c|}{ $IEF\sim GDP^{\gamma}$}  \\ \hline
year&$\gamma$&$\Delta \gamma$&$\Delta \gamma/\gamma$&$ R^2$\\
\hline
1996 &0.0940 &0.0117 &0.1248 &0.3255\\
1997 &0.0935 &0.0109 &0.1163 &0.3439\\
1998 &0.0994 &0.0113 &0.1140 &0.3435\\
1999 &0.0956 &0.0112 &0.1166 &0.3261\\
2000 &0.0915 &0.0099 &0.1086 &0.3583\\
2001 &0.0870 &0.0098 &0.1131 &0.3472\\
2002 &0.0824 &0.0101 &0.1224 &0.3107\\
2003 &0.0802 &0.0075 &0.0940 &0.4332\\
2004 &0.0773 &0.0073 &0.0947 &0.4313\\
2005 &0.0728 &0.0070 &0.0956 &0.4267\\
2006 &0.0662 &0.0064 &0.0961 &0.4208\\
2007 &0.0670 &0.0062 &0.0922 &0.4414\\
\hline
 \end{tabular} 
   \caption{
   Yearly evolution of the power law exponent ($\gamma$) between the IEF and GDP, the standard error
($\Delta \gamma$), the relative error ($\Delta \gamma/\gamma$), and the efficiency ($R^2$) of the regression. 
There are 3 periods to be noticed 
in which the exponent adopts different behaviours. For the years 1996 to 2000, the exponent  has an average value $ \simeq$ 0.0948 and  remains
stable  ($\simeq 0.09$) for about 5 years. The second phase  spreads over the years 2001 to 2005, is a transitional period during which the value of
the exponent  falls down. It ends up stabilising around 0.0666 on the third and latest period (2006-2007). 
Notice that the regression coefficient ($R^2$) is not very high. } 
   \label{TablegammaIEFGDP}
\end{center} \end{table}

In Table \ref{TablegammaIEFGDP}, 
we report  the power  law exponent ($\gamma$) between the IEF and GDP, the standard error
($\Delta \gamma$), the relative error ($\Delta \gamma/\gamma$), and  the  ($R^2$)   regression  coefficient. There are 3 periods to be noticed 
in which the exponent adopts different behaviours. For the years 1996 to 2000, the exponent  has an average value $ \simeq$ 0.0948 and  remains
stable  for about 5 years. The second phase, which is spread over the years 2001 to 2005, is a transitional period during which the value of
the exponent  falls down. It ends up stabilising around 0.0666 on the third and latest period (2006-2007). 
Notice that the regression coefficient is not very high: $R^2 \sim 0.376$.
  
  \clearpage
  
\section{Conclusions}\label{Conclusions} 
Let us recall the Research Questions: can one find an empirical law for describing the economic freedom (EF)  of nations through the main measure indices, i.e. the Economic Freedom of the World (EFW) index \cite{[23]} and the  Index of Economic Freedom (IEF) \cite{[24]} ?
What simple empirical laws can be found through a simple analysis  of rank-size laws ?
Are such laws of interest for discussing the main determinant, according to the literature,  i.e. each country GDP?

We have taken  some   data pertaining to   the 1997-2007 period, that is before 2008, thus before a recent "financial crisis", in order not to involve "multiple exaggerated developments" \cite{Wigmore}, but  nevertheless in order to include a drastic turning point, Sept. 11, 2001,  following another  geo-economico-political event,  the fall of the Berlin wall. 
We have pointed out that the study of EF should develop over two distinct periods, at this time, mainly because the Index's 2008 definition of economic freedom has been modified. 
In so doing we have  selected data, leading to 138 countries  examined  over a period extending from 2000 to 2006, thus 2 sets of 862 data points.

We have found that   the rank distributions obey either an exponential or a power law or a mixed behaviour.   The EFW-rank relationship is exponential for countries of high rank ($\ge  20$), but  log-log plots point to a behaviour close to a power law when considering the whole sample. 
In contrast,  the IEF  overall ranking has   an exponential behaviour. Interestingly,  IEF rank-size rule log-log plots point to  the existence of a transitional point between two different power laws, i.e., near rank 10. 

Besides, the IEF appears to be "more conservative" than the EFW index. 

Moreover, when searching for  (analytical law)  correlations between the country GDP   and  either EF indices\footnote{We have not looked for regressions between these macroeconomic variables and the various "pillars" of the indices, - the literature is already abundant.},  we have distinguished regional aspects, i.e. defining 6 continents.
 We find that  the EFW index relationship to  country GDP   is characterised by a power law, with a rather stable exponent ($\gamma \simeq 0.674$) as a function of time. In contrast,   the IEF relationship with  the country's gross domestic product  points to a downward evolutive  power law parameter as a function of time. Markedly the two studied indices provide different aspects of EF.
 
   In so doing, we add numerical considerations  to the literature, as should be somewhat expected by econophysics research, - for this Special Issue, but presenting to others a different perspective. The rank-size law approach seems original for the present topics. It brings some information on the ''statistical universality'' of the EF during the consider time interval. Thus we expect to open  gates for rigorous approaches, i.e. stressing objectiveness in the modelling, rather than ideological bases.

Thereafter, suggestions for further research can be listed:  among others, one could consider other time intervals; for example including the 2008 financial crisis, and nowadays considering the COVID-19 pandemic! This is left for our expected paper II. 
On the other hand, It would be nice to have  more "economic considerations" and "historical considerations" looking in more details at each pillar separately.
 For example,  one could consider some renormalisation of the indices, taking into account,  size (and type)  of governments, size of country populations, inflation rates, foreign direct investments, health burden, etc., on one hand, and on the other hand, migration factors, religious factors, education levels, trade union strengths,  pandemic constraints,  local climate, etc. 
  Quite a numerical challenge to econophysicists tough.

\vskip1.5cm
{\bf Author Contributions:} The authors have contributed equally. All authors have read and agreed to submitting this  version of the manuscript.
\vskip1.5cm
{\bf Funding}: MA has been partially supported by a grant of the Romanian National Authority for Scientific Research and Innovation, CNDS-UEFISCDI, project number PN-III-P4-IDPCCF-2016-0084.
 
\vskip1.5cm
{\bf  Conflicts of Interest}:  The authors declare no conflict of interest.

\newpage

\end{document}